\documentstyle[twoside,fleqn,espcrc2,psfig]{article}

\def\Journal#1#2#3#4{{#1} {\bf #2}, #3 (#4)}

\def\NPB{{\em Nucl. Phys.} B}
\def\PLB{{\em Phys. Lett.}  B}
\def\PRL{\em Phys. Rev. Lett.}
\def\PRD{{\em Phys. Rev.} D}
\def\ZPC{{\em Z. Phys.} C}
\def\ZPA{{\em Z. Phys.} A}
\def\NPA{{\em Nucl. Phys.} A}
\def\JPG{{\em J. Phys.} G}
\def\nn{\nonumber}

\def\ra{\rightarrow}
\def\vk{{\bf k}_{\perp}}
\def\al{\alpha_s}
\def\be{\begin{equation}}
\def\ee{\end{equation}}
\def\bea{\begin{eqnarray}}
\def\eea{\end{eqnarray}}

\def\kev{\,{\rm keV}}
\def\mev{\,{\rm MeV}}
\def\gev{\,{\rm GeV}}

\newcommand{\AmS}{{\protect\the\textfont2
  A\kern-.1667em\lower.5ex\hbox{M}\kern-.125emS}}

\title{\bf On hard exclusive reactions in the time-like region}

\author{P.\ Kroll\address{Fachbereich Physik, Universit\"at Wuppertal\\
        Gau\ss strasse 20, D-42097 Wuppertal, Germany}
         } 
       
\begin{document}
\thispagestyle{empty}
\begin{flushright}
WU-B 96-38\\
November  1996 \\[1.3cm]
\end{flushright}            
\renewcommand{\thefootnote}{\alph{footnote}}
\begin{center}
{\Large\bf On hard exclusive reactions in the time-like region 
\footnote {Invited contribution presented at the Fourth Biennal
Conference on Low-Energy Antiproton Physics, LEAP96, Dinkelsb\"{u}hl
(August 1996) }} \\
\vspace*{1.0 cm}
\end{center}
\begin{center}
{\large P. Kroll}\\
\vspace*{0.5 cm}
Fachbereich Physik, Universit\"{a}t Wuppertal, \\
D-42097 Wuppertal, Germany\\[0.3 cm]
\end{center}
\setcounter{page}{0}
\newpage
\setcounter{footnote}{0}
\renewcommand{\thefootnote}{\arabic{footnote}}
\begin{abstract}
The proton form factor, two-photon annihilations into $p\bar{p}$ 
as well as exclusive charmonium decays are critically examined. 
It will be argued that the standard perturbative QCD analysis of these
reactions fails, i.~e.~the need for additional contributions can convincingly
be demonstrated. Possible dynamical mechanisms such as colour-octet
admixtures to the charmonium states or diquarks inside baryons, will 
be discussed and compared to the data. 
\end{abstract}

\maketitle

\section{INTRODUCTION}

At large Mandelstam $s$ and large momentum transfer the hard
scattering approach (HSA) \cite{lep:80} provides a scheme to calculate
exclusive processes. Observables are described as convolutions of
hadronic wave functions which embody soft non-perturbative physics,
and hard scattering amplitudes $T_H$ to be calculated from
perturbative QCD. In most cases only the contribution from the
lowest-order pQCD approach in the collinear approximation using
valence Fock states only (termed the standard HSA) has been worked out. 
Applications of the standard HSA to space-like
exclusive reactions, as for instance the magnetic form factor of the
nucleon, the pion form factor or Compton scattering off protons
revealed that the results are only in fair agreement with experiment
if strongly end-point region (where one of the quark momentum
fractions, $x$, tends to zero) concentrated hadronic wave functions are
used. As has been pointed out by several authors (e.g.\ \cite{isg:89}),  
the results obtained from such wave functions are dominated by contributions
>from the end-point regions where perturbative QCD cannot
readily be applied. Hence, despite the agreement with experiment, the
predictions of the standard HSA are theoretically inconsistent for 
such wave functions. It should also be stressed that the large
momentum transfer behaviour of the helicity-flip controlled
Pauli form factor of the proton remains unexplained within the standard HSA. 

Applications of the HSA to time-like exclusive processes fail in most
cases (e.g.\ $G_M$, $F_{\pi}$, $\gamma\gamma\to p\bar{p}$). 
The predictions for the integrated  $\gamma\gamma\to \pi\pi$
cross-section ($|\cos{\theta}| \leq 0.6$) are in fair
agreement with the data whereas the predictions for the angular
distribution fails. Exclusive charmonium decays constitute another
class of time-like reactions. If the end-point concentrated wave
functions are employed again, the standard HSA provides results 
in fair agreement with the data in many cases. Noteworthy are the 
failures for the decays of the $\eta_c$ and the $\chi_{c0}$ into 
$p\bar{p}$. The standard HSA predicts zero decay widths for these 
reactions while experimentally the decay widths and
the branching ratios are of similar magnitude as those of 
the other charmonium decays into $p\bar{p}$. The reason for this
failure is obvious:
the perturbative mechanism produces only $p\bar{p}$ pairs with opposite
helicities while the quantum numbers of the $\eta_c$ and the
$\chi_{c0}$ require pairs with the same helicities. It
should also be noted that in most calculations of exclusive
charmonium decays \cite{dun:80} $\al$ values of the 
order of $0.2 - 0.3$ are employed. Such values do not match with $\al$ 
evaluated at the charm quark mass, the characteristic scale for these 
decays ($\al (m_c=1.5 \gev) = 0.37$ in one-loop approximation with 
$\Lambda_{QCD}=200 \mev$). Since high powers of $\al$ are involved in 
charmonium decays a large factor of uncertainty is hidden in the predictions.   

In this talk I am going to discuss higher Fock state corrections to 
the standard HSA. Constraining the pion wave function from
the recent precise data on the $\pi\gamma$ transition form factor
\cite{cleo:95}, one observes an order-of-magnitude 
discrepancy between data and HSA predictions for charmonium decays
into two pions. In \cite{bol:96} contributions from the 
$c\bar{c}g$ Fock state are suggested as the solution of this puzzle.
In order to cure the failure of the standard HSA for reactions 
involving protons a variant of the HSA has been proposed 
\cite{kro:93a}${-}$\cite{jak:93} in which the proton is viewed as
being composed of quarks and diquarks. The latter objects constitute
a particle model for higher Fock state contributions. 

\section{THE $\pi\gamma$ FORM FACTOR}

The apparent success of the end-point concentrated wave
functions, in spite of the theoretical inconsistencies, prevented progress
in understanding hard exclusive reactions for some time. Recently the 
situation has changed with the advent of the CLEO data on the
$\pi\gamma$ transition form factor $F_{\pi\gamma}$ \cite{cleo:95}. The
leading twist result for that form factor, including
$\al$-corrections, reads \cite{lep:80}
\bea
\label{leadtwisteq}
F_{\pi\gamma}(Q^2)&=&\frac{\sqrt 2}{3}\,\langle
                 x^{-1}\rangle\frac{f_\pi}{Q^2}\;\times\\
             &&[\,1+\frac{\alpha_s(\mu_R)}{2\pi}
               K(Q^2,\mu_R) + {\cal O}(\alpha_s^2)\,]. \nn
\eea 
The function $K$ has been calculated by Braaten \cite{bra:83}.
$f_{\pi}$ is the usual pion decay constant
(130.7 \mev) and $\langle x^{-1}\rangle$ is the $1/x$ moment of the
pion distribution amplitude (DA), $\phi$, which represents the
light-cone wave function of the pion integrated over transverse
quark momenta, $\vk$, up to a factorization scale, $\mu_F$, of order $Q$. 
The DA can be expanded upon Gegenbauer polynomials, $C_n^{3/2}$, the 
eigenfunctions of the evolution kernel for mesons \cite{lep:80}
\bea
\label{evoleq}
&&\phi_{\pi}(x,\mu_F)=\phi_{AS}(x)\left[1+ \right.  \\
    && \sum^\infty_{n=2,4,...}B_n(\mu_0)\left(
\frac{\al\left(\mu_F\right)}{\al\left(\mu_0\right)}\right)
^{\gamma_n}\,C_n^{3/2}(2x-1)] \nn
\eea
where the asymptotic DA is $6x(1-x)$. The process-in\-depen\-dent
ex\-pan\-sion coefficients $B_n$ embody the soft physics; they are not
calculable at present. The $\gamma_n$ are the anomalous
dimensions. $\mu_0$ is a typical hadronic scale, actually $\mu_0=0.5 \gev$.
Any DA evolves into the asymptotic DA for $\ln{Q^2}\to\infty$. Hence, 
the limiting behaviour of the transition form factor is
\be
\label{asy}
F_{\pi\gamma} \longrightarrow \sqrt{2} f_{\pi} /Q^2
\ee
which is a parameter-free QCD prediction. As comparison with the
CLEO data \cite{cleo:95} reveals, the limiting
behaviour is approached from below. At 8 \gev$^2$ the data only
deviate by about $10 - 15 \%$ from (\ref{asy}). The leading twist
result (\ref{leadtwisteq})  without (with) $\al$-corrections nicely
fits the CLEO data for $B_2(\mu_0)=-0.39\;(-0.17)\, \pm 0.05 $ and $B_n=0$
($n\geq 4$) \cite{jak:96,kro:96}. I.e.\ the required DA is narrower 
than the asymptotic one in the momentum transfer region of a few 
\gev$^2$. The frequently used Chernyak-Zhitnitsky DA
\cite{che:82}, defined by $B_2(\mu_0) =2/3$, $B_n=0$ ($n\geq 4$), is in clear
conflict with the data and should, therefore, be discarded.

Recently a modified HSA has been proposed by Botts, Li and Sterman
\cite{bot:89} in which transverse degrees of freedom as well as
Sudakov suppressions are taken into account. This approach has the
advantage of strongly suppressed end-point regions. Hence, the
perturbative contributions can be calculated self-consistently. Using
a Gaussian for the $\vk$-dependence of the pion wave function
\bea
\label{gaussian}
\Psi_{\pi}\left(x,\vk;\mu_F\right)&=&\frac{f_\pi}{2\sqrt 6}\,
                                           \phi_{\pi}(x,\mu_F)\\
            &&\times
            N\exp{\left(-a^2_{\pi}\frac{k_{\perp}^2}{x(1-x)}\right)} \nn
\eea
where $N=16\pi^2 a^2_{\pi}/(x(1-x))$ and, for a DA with $B_n=0$ for 
$n\geq 4$, $a_{\pi}=1/(\pi f_{\pi} \sqrt{8(1+B_2)}$ which
automatically satisfies the $\pi^0\to \gamma\gamma$ constraint 
\cite{bro:83}, ond finds perfect agreement with the CLEO data 
for $B_2(\mu_0) =-0.006 \pm 0.014$ \cite{jak:96,kro:96}, i.e.\ the 
asymptotic wave function works very well if the modified HSA is used.

\section{PION DECAYS OF CHARMONIUM}      

In view of the results for $F_{\pi\gamma}$ a fresh analysis of the
decays $\chi_{cJ}\to \pi\pi$ is in order. Using the information on 
the $\pi$ wave function obtained from the analysis of $F_{\pi\gamma}$,
one finds the following values for the partial widths 
\be
\label{shsa}
\Gamma (\chi_{c0(2)}\to\pi^+\pi^-)\, =\, 0.872\; (0.011)\, \kev 
\ee
within the standard HSA \cite{bol:96}. As usual the renormalization 
and the factorization scales are identified in that calculation and 
put equal to the $c$-quark mass. The parameter describing the 
$\chi_{cJ}$ state is the derivative $R'_P(0)$ of the
non-relativistic $c\bar{c}$ wave function at the origin (in coordinate
space) appropriate for the dominant Fock state of the $\chi_{cJ}$, a
$c\bar{c}$ pair in a colour-singlet state with quantum numbers
${}^{2S+1}L_J={}^3P_J$. $m_c=1.5\;\gev$  and, of course, $B_2(\mu_0) =-0.39$ are
chosen as well as $R'_P(0)=0.22\gev^{5/2}$ which is consistent with 
a global fit of charmonium parameters \cite{man:95} as well as with results
for charmonium radii from potential models \cite{buc:81}.
 
In \cite{bol:96} also the modified HSA is used to calculate the
$\chi_{cJ}\to \pi\pi$ decay widths. Taking  $B_2=0$
and the other parameters as quoted above, one finds
\be
\label{mhsa}
\Gamma (\chi_{c0(2)}\to\pi^+\pi^-)\, =\, 8.22\; (0.41)\, \kev. 
\ee
For comparison the experimental values as quoted in \cite{pdg} and
reported in a recent paper of the BES collaboration \cite{bes} are
\bea
\label{dat}
\Gamma (\chi_{c0}\to\pi^+\pi^-)&=& 105\; \pm 30\phantom{.3} \kev \;({\rm PDG}),\nn \\ 
                               && 62.3\pm 17.3 \kev \;({\rm BES}), \nn \\ 
\Gamma (\chi_{c2}\to\pi^+\pi^-)& =& 3.8\;\;\pm 2.0\phantom{3}
                                                      \kev \;({\rm PDG}),\nn \\
                               &&  3.04 \pm 0.73 \kev\; ({\rm BES}). 
\eea
One notes that both the theoretical results, (\ref{shsa}) and
(\ref{mhsa}), fail by at least an order of magnitude. To assess the
uncertainties of the theoretical results one may vary the parameters,
$m_c$, $B_2$ and $\Lambda_{QCD}$. However, even if the parameters are
pushed to their extreme values the predicted rates are well below
data. Thus, one has to conclude that calculations based on the
assumption that the $\chi_{cJ}$ is a pure $c\bar{c}$ state, are not
sufficient to explain the observed rates. The necessary corrections
would have to be larger than the leading terms. A new mechanism is
therefore called for. 

\begin{figure}[t]
\unitlength 1mm
 \begin{picture}(70,40)
   \put( 2.1, 6){\psfig{figure=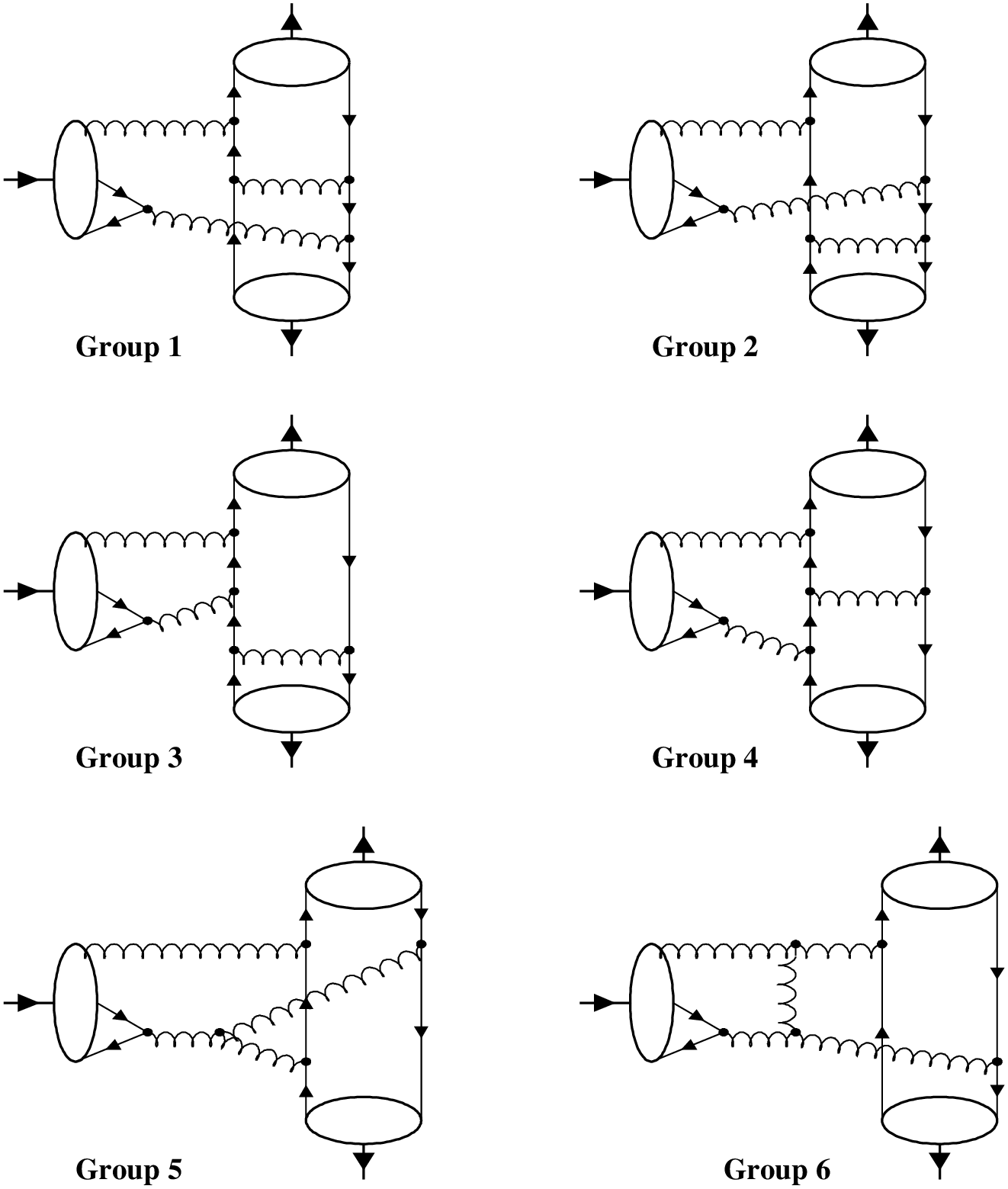,%
       bbllx=10pt,bblly=335pt,bburx=580pt,bbury=810pt,%
       height=3.0cm,clip=} }
   \put(40.0, 6){\psfig{figure=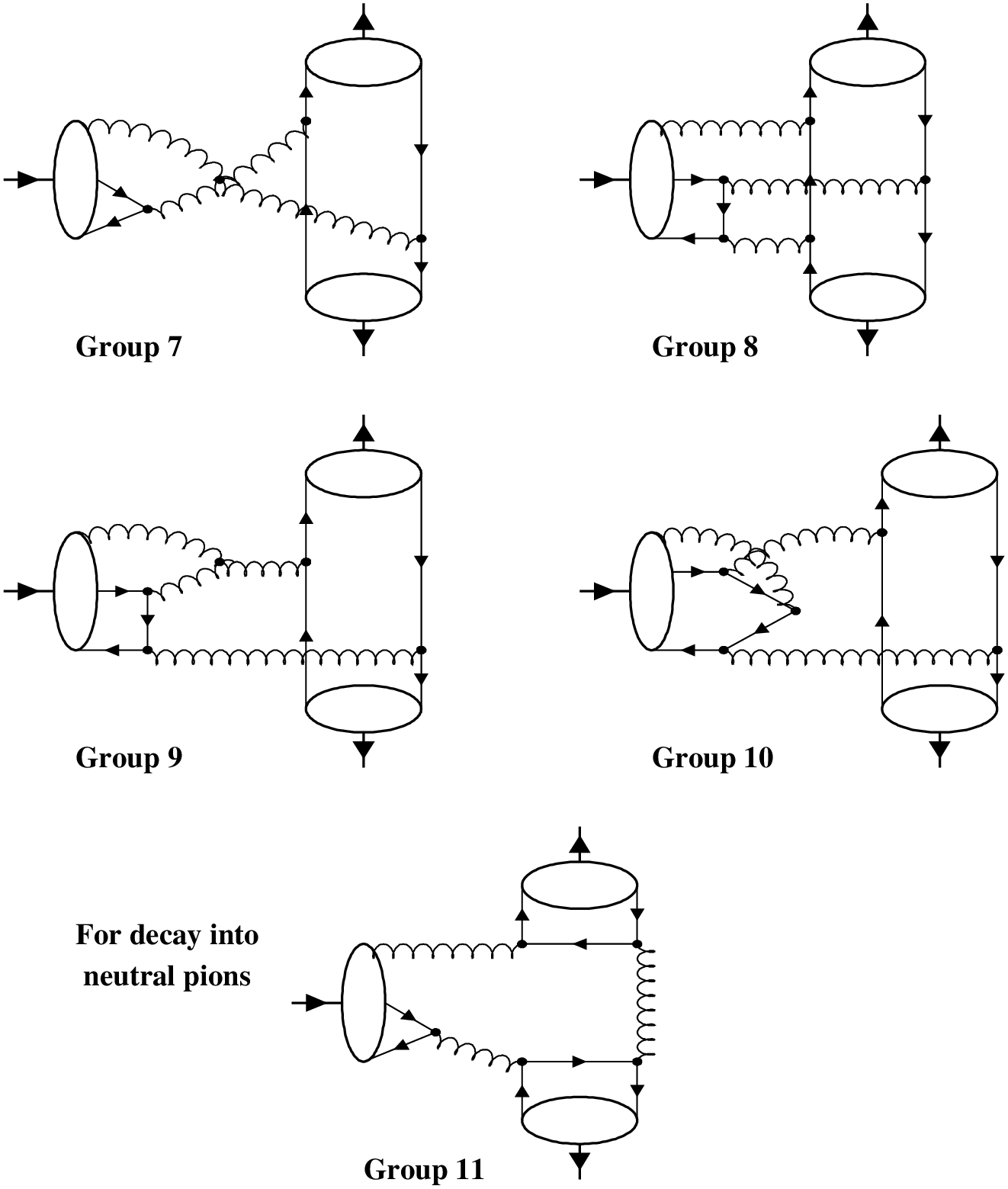,%
       bbllx=10pt,bblly=335pt,bburx=580pt,bbury=810pt,%
       height=3.0cm,clip=} }
 \end{picture}
\caption{Representatives of the various groups of colour-octet decay graphs.
 \label{fig:col}}
 \vspace{-0.3cm}
\end{figure}

Recently, the importance of higher Fock states in understanding the
production and the {\em inclusive} decays of charmonium has been
pointed out \cite{bod:95}. It is therefore tempting to assume the inclusion of
contributions from the $|c\bar{c}_8 (^3S_1)g\rangle$ Fock state to
{\em exclusive} $\chi_{cJ}$ decays as the solution to the failure of
the HSA. The usual higher Fock state suppression by powers of $1/Q^2$
\cite{lep:80} where $Q=m_c$ in the present case, does not appear as a
simple dimensional argument reveals; both the contributions to the
decay amplitude, the colour-singlet and the octet one, behave as
$1/m_c^3$. In \cite{bol:96} the colour-octet contributions to the
exclusive $\chi_{cJ}$ decays are estimated by calculating the hard
scattering amplitude from the set of Feynman graphs shown in Fig.\
\ref{fig:col} and convoluting it with the asymptotic pion wave function.
The colour-octet and singlet contributions are to be added coherently. The
$\chi_{cJ}\to \pi\pi$ decay widths are given in terms of a single
non-perturbative parameter $\kappa$ which approximately accounts for
the soft physics in the colour-octet contribution. A fit ot the data
\cite{pdg,bes} yields $\kappa=0.16 \gev^2$ and the widths
\be
\label{so}
\Gamma (\chi_{c0(2)}\to\pi^+\pi^-)\, =\, 49.85\; (3.54)\,  \kev. 
\ee
Comparison with (\ref{dat}) reveals that the inclusion of the
colour-octet mechanism brings predictions and data in generally good
agreement. The value found for the parameter $\kappa$ has a reasonable
interpretation in terms of charmonium properties and the mean transverse 
momentum of the quarks inside the pions. Thus it seems that the 
colour-octet mechanism leads to a satisfactorily explanation of the 
decay rates of the $\chi_{cJ}$ into two pions. Of course, that 
mechanism has to pass more tests in exclusive reactions before this 
issue can be considered as being settled.  

\section{REACTIONS INVOLVING PROTONS}

The standard HSA runs into many difficulties with these exclusive reactions
as mentioned in the introduction. In a series of papers 
\cite{kro:93a}${-}$\cite{jak:93} a variant of the HSA has been
proposed in which baryons are assumed to be composed
of quarks and diquarks. A diquark, being a cluster of two valence
quarks and a certain amount of glue and sea quarks pairs, is regarded
as a quasi-elementary constituent. In so far, a quark-diquark state
represents an (unspecified) superposition of higher Fock states
\bea
\label{fock}
|B,\lambda>& = &\Psi_{qD}^B(x,\vk) |qD> \\
           &=& \Psi^B_{qqq} |qqq> + \Psi^B_{qqqg} |qqqg> + .....\nn
\eea 
In the diquark model spin $0$ ($S$) and spin $1$ ($V$)
colour-antitriplet diquarks are considered. Assuming zero relative orbital 
angular momentum between quark and diquark and taking advantage of 
the collinear approximation, the valence Fock state of a proton 
with helicity $\lambda$ and momentum $p$ can be written in a 
covariant fashion (omitting colour indices)
\bea
\label{pwf}
|P;p,\lambda\rangle&=&f_S\Phi_S(x)\,B_S\, u(p,\lambda) \\
            &+&\frac{f_V}{\sqrt{3}}\Phi_V(x) B_V
              (\gamma^{\alpha}+p^{\alpha}/m)\gamma_5 u(p,\lambda) \nn
\eea
where $u$ is the proton's spinor. The two terms in (\ref{pwf})
represent configurations consisting of a quark and either a scalar or a 
vector diquark, respectively. The couplings of the diquarks 
with the quarks in a spin-isospin 1/2 baryon lead to the flavour functions 
\bea
\label{fwf}
B_S&=&u\, S_{[u,d]}   \nn\\ 
B_V&=& [ u V_{\{u,d\}} -\sqrt{2} d\, V_{\{u,u\}}]/\sqrt{3}\, .
\eea
In the diquark model the following DAs have been proven to work
satisfactorily well in many applications \cite{kro:93a}${-}$\cite{jak:93}:
\bea
\label{a10}
\Phi_S(x)&\hspace{-0.3cm}=&\hspace{-0.3cm}N_S x (1-x)^3 
                                 \exp{\left[-b^2 (\frac{m^2_q}{x} 
                                  + \frac{m^2_S}{1-x})\right]}\nn \\
\Phi_V(x)&\hspace{-0.3cm}=&\hspace{-0.3cm}N_V x (1-x)^3 
             (1+5.8\,x - 12.5\,x^2) \nn\\
         &&        \times\exp{\left[-b^2 (\frac{m^2_q}{x} +
                                   \frac{m^2_V}{1-x})\right]}.  
\eea
These DAs are suitable adaptions of a meson DA obtained by
transforming the harmonic oscillator wave function to the
light cone. The constants $N$ are fixed through the
normalization convention ($N_S = 25.97$ and $N_V = 22.92$). 
The exponentials in (\ref{a10}) guarantee a strong suppression of the 
end-point regions. The masses appearing in the exponentials are 
constituent masses since they enter through a rest frame wave 
function. For $u$ and $d$ quarks $350\mev$ and for the diquarks 
$580\mev$ are appropriate mass values. It is to be stressed that
the quark and diquark masses only appear in the DAs (\ref{a10}); in
the hard scattering kinematics they are neglected. 
The transverse size parameter $b$ is fixed from the assumption of a
Gaussian transverse momentum dependence of the full wave function and
the requirement of a value of $600\mev$ for the mean transverse
momentum (actually $b = 0.498\gev^{-1}$). The constituent masses 
and the transverse size parameter are not considered as
free parameters since the final results only depend on them mildly.
 
Diquark-gluon and diquark-photon vertices appear
in the Feyman graphs contributing to the hard scattering amplitude
of a given process. Following standard prescriptions, these vertices
are defined as 
\bea
\label{vert}
\mbox{SgS}:&& i\,g_s t^{a}\,(p_1+p_2)_{\mu} \nn\\
\mbox{VgV}:&& -i\,g_{s}t^{a}\, 
\Big\{
 g_{\alpha\beta}(p_1+p_2)_{\mu} \\
&&- g_{\beta\mu}\left[(1+\kappa)\,p_2-\kappa\, p_1\right]_{\alpha} \nn\\
&& - g_{\mu\alpha} \left[(1+\kappa)\,p_1-\kappa\, p_2\right]_{\beta} 
\Big\} \nn
\eea
where $g_s=\sqrt{4\pi\alpha_s}$ is the QCD coupling constant.
$\kappa$ is the anomalous magnetic moment of the vector diquark and 
$t^a=\lambda^a/2$ the Gell-Mann colour matrix. For the coupling of 
photons to diquarks one has to replace $g_s t^a$ by $-\sqrt{4\pi\alpha} e_D$ 
where $\alpha$ is the fine structure constant and $e_D$ is the electrical 
charge of the diquark in units of the elementary charge. The couplings 
$DgD$ are supplemented by appropriate contact terms required by 
gauge invariance.

The composite nature of the diquarks is taken into 
account by phenomenological vertex functions. Advice for the parameterization 
of the 3-point functions (diquark form factors) is 
obtained from the requirement that asymptotically the diquark 
model evolves into the standard HSA. In so far the standard HSA and
the diquark model do not oppose each other, they are not alternatives
but rather complements. Interpolating smoothly  between 
the required asymptotic behaviour and the conventional value of 1 at $Q^{2}=0$, 
the diquark form factors are actually parametrized as
\bea
\label{fs3}
\vspace*{-0.5cm}
F_{S}^{(3)}(Q^{2})&=&\frac{Q_{S}^{2}}{Q_{S}^{2}+Q^{2}} \nn \\
F_{V}^{(3)}(Q^{2})&=&\left(\frac{Q_{V}^{2}}{Q_{V}^{2}+Q^{2}}\right)^{2}
\eea
in the space-like region. The asymptotic behaviour of the diquark 
form factors and the connection to 
the hard scattering model is discussed in more detail in \cite{kro:91,jak:93}.  
In accordance with the required asymptotic behaviour the $n$-point
functions for $n\geq 4$ are parametrized as
\bea
\label{fsn}
\vspace*{-0.8cm}
F_{S}^{(n)}(Q^{2})&=&a_{S}F_{S}^{(3)}(Q^{2})\,, \\ 
F_{V}^{(n)}(Q^{2})&=&
\left(a_{V}\frac{Q_{V}^{2}}{Q_{V}^{2}+Q^{2}}\right)^{n-3}F_{V}^{(3)}(Q^{2}).
                                                      \nn
\eea
The constants $a_{S,V}$ are strength parameters. Indeed, since the diquarks in 
intermediate states are rather far off-shell one has to consider 
the possibility of diquark excitation and break-up. Both these possibilities 
would likely lead to inelastic reactions. Therefore, these
possibilities are not considered in the diquark model explicitly but 
taken into account by the strength parameters. Since in most 
cases the contributions from the n-point functions for $n\geq 4$ only 
provide small corrections to the final results that recipe is 
sufficiently accurate.

The relations (\ref{fs3},\ref{fsn}) represent effective
parameterizations valid at large space-like $Q^2$. It is not possible
to continue these parameterizations to the time-like region in a
unique way since the exact dynamics of the diquark system is unknown. 
A suitable continuation to the time-like region is defined by the
following prescription: $Q^2$ is replaced by $-s$ in
(\ref{fs3},\ref{fsn}) which ensures the correct asymptotic behaviour and,
in order to avoid the appearance of unphysical poles, the diquark form
factors are kept constant once their absolute values have reached a
certain value ($c_0=1.3$) \cite{kro:93a}.

\begin{figure}[t]
\vspace{9pt}
 \psfig{figure=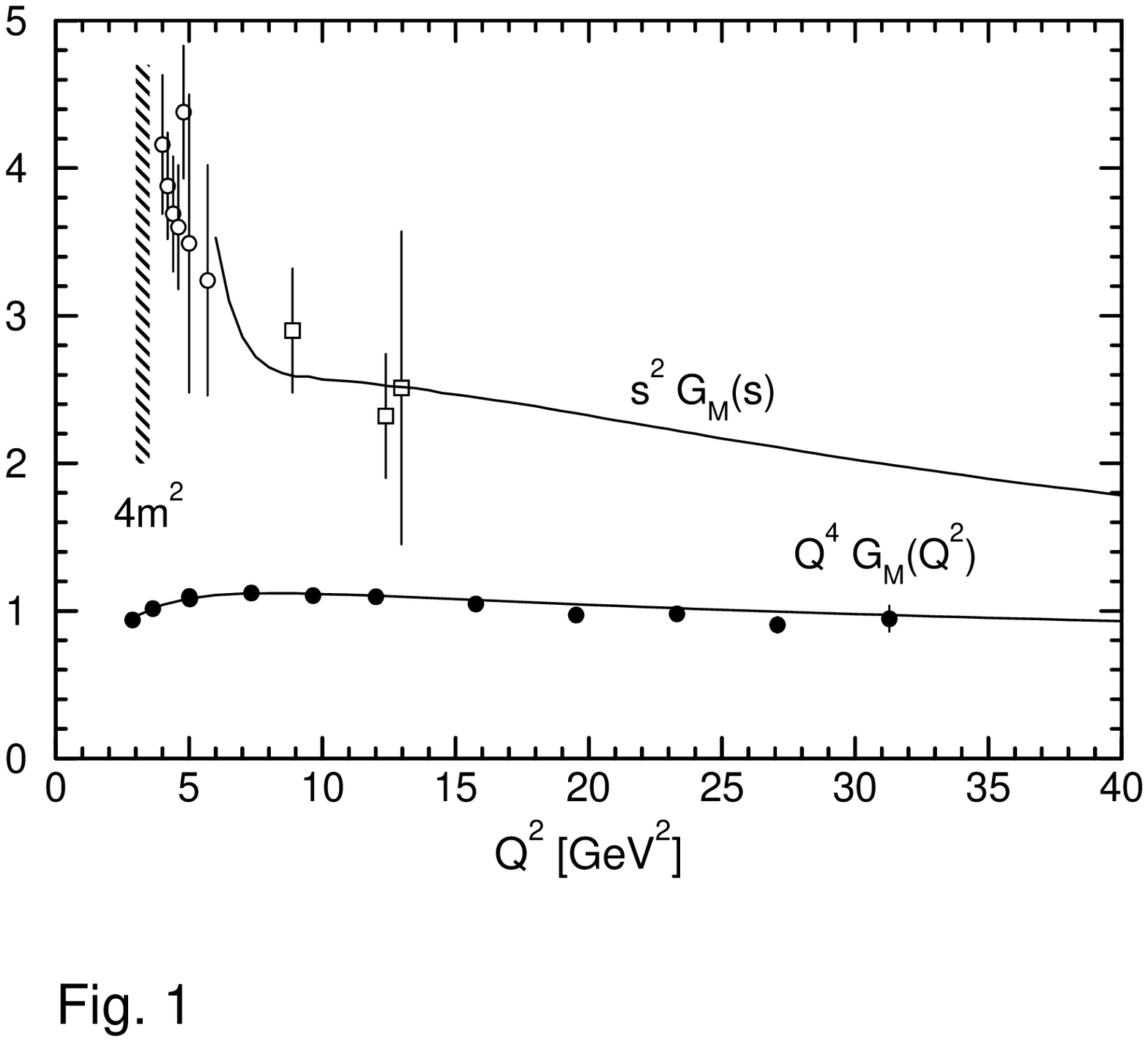,%
       bbllx=45pt,bblly=105pt,bburx=535pt,bbury=485pt,%
       height=4.3cm,clip=}
\vspace*{-10mm}
\caption[]{The magnetic form factor of the proton in the time-like and
  space-like (at $Q^2=-s$) regions. The solid line represents the
  predictions of the diquark model \cite{kro:93a}. The time-like data 
  ($\circ$, $\Box$) are taken from \cite{arm:93,bis:90}, the
  space-like data ($\bullet$) from \cite{arn:86}.
\label{fig:ff}}
\vspace*{-0.3cm} 
\end{figure}

\begin{figure}[t]
 \vspace{9pt}
 \psfig{figure=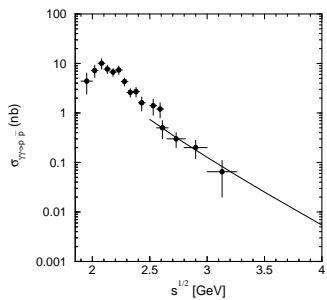,%
        bbllx=70pt,bblly=230pt,bburx=465pt,bbury=575pt,%
        height=4cm,clip=}
 \vspace*{-10mm}
\caption[]{The integrated $\gamma\gamma\ra p\bar{p}$ cross section
  ($\mid\cos{\theta}\mid\le 0.6$). The solid line represents the diquark
  model prediction \cite{kro:93a}. Data are taken from CLEO \cite{cleo}.} 
\label{fig:gg}
\vspace*{-0.5cm} 
\end{figure}

The analysis of electromagnetic nucleon form factors constitutes the simplest
application of the diquark model and the most obvious place to fix the
various parameters of the model. The Dirac and Pauli form factors (the
necessary helicity flips are provided by the vector diquarks) of
the nucleon are evaluated by convoluting the DAs (\ref{a10}) with the
respective hard scattering amplitude \cite{lep:80}. The parameters are
determined from a best fit to the data in the space-like region. The 
following set of parameters
\[
\hspace{-2mm}
\begin{array}{ccc}
 f_S= 73.85\,\mbox{MeV},& Q_S^2=3.22 \,\mbox{GeV}^2, & a_S=0.15,  \\
 f_V=127.7\,\mbox{MeV},& Q^2_V=1.50\,\mbox{GeV}^2, & a_V=0.05,  \\
 \phantom{f_V=127.7\,\mbox{MeV},}&\kappa=1.39
\end{array}
\]
provides a good fit of the data \cite{jak:93}. The 
parameters $Q_S$ and $Q_V$, controlling the size of the diquarks,
are in agreement with the higher-twist effects observed in the structure 
functions of deep inelastic lepton-hadron scattering if these 
effects are modelled as lepton-diquark elastic scattering \cite{vir:91}. 
The predictions for the magnetic form factor in both the
space-like and the time-like regions, are compared to the data 
\cite{arm:93,bis:90,arn:86} in Fig.\ \ref{fig:ff}.
%
Two-photon annihilation into $p\bar{p}$ pairs has also been
investigated within the diquark model. The prediction for the
integrated $\gamma\gamma\ra p\bar{p}$ cross section is compared to the
CLEO data \cite{cleo} in Fig.\ \ref{fig:gg}. At large energies the agreement
between prediction and experiment is good. The prediction for the 
angular distribution is in agreement with the CLEO data too. 

The diquark model also allows to investigate the decay process
$\eta_c\to p \bar{p}$. A calculation along the same lines as for the
other two time-like processes, using the same DAs and the same set of
parameters, leads to a decay width of $3.88$ keV \cite{kro:93a} which is in fair
agreement with the data \cite{pdg}. Note that in the pure quark HSA a zero
width is obtained.

\section{SUMMARY}

The study of hard exclusive reactions is an interesting and
challenging subject. The standard HSA, i.e.\ the valence Fock state
contribution in collinear approximation to lowest
order perturbative QCD, while asymptotically correct (at
least for form factors), does not lead to a consistent description of
the data. In many cases the predicted perturbative contribution to
particular exclusive reactions are much smaller then the data. The
observed spin effects do not find a comforting explanation. In some
reactions agreement between prediction and experiment is found
although at the expense of dominant contributions from the soft
end-point regions rendering the perturbative analysis inconsistent.

In view of these observations it seems that higher
Fock state contributions have to be included in the analysis. However,
not much is known about them as yet. We are lacking systematic
investigations of such contributions to exclusive reactions. A few
examples of such contributions have been discussed in this talk,
namely the colour octet model for exclusive charmonium decays and the
diquark model. More work is needed.

\end{document}